# NEUTRON STARS WITH A QUARK CORE.  II.  BASIC INTEGRAL AND STRUCTURAL PARAMETERS

G. B. Alaverdyan[1], A. R. Harutyunyan[2], and Yu. L. Vartanyan[3]

*A broad sample of computed realistic equations of state of superdense matter with a quark phase transition is used to construct a series of models of neutron stars with a strange quark core. The integral characteristics of the stellar configurations are obtained: gravitational mass, rest mass, radius, relativistic moment of inertia, and red shift from the star's surface, as well as the mass and radius of the quark core within the allowable range of values for the central pressure. The parameters of some of the characteristic configurations of the calculated series are also given and these are studied in detail. It is found that a new additional region of stability for neutron stars with strange quark cores may exist for some models of the equation of state.*

Keywords:  *stars:neutron:quark core*

## 1. Introduction

The properties of neutron stars have a functional dependence on the form of the equation of state (ES) of matter over an extremely wide range of densities extending from 7.86 g/cm$^3$ to on the order of $10^{15}$ g/cm$^3$.  A quark degree of freedom may play a controlling role at supranuclear densities.  In part I of this paper [1] we have constructed a broad set of realistic equations of state for superdense matter which yield a first order phase transition from a state in which quarks are contained inside baryons to a continuous quark-electron plasma state.

In this paper a series of models of layered neutron stars with a strange quark core are calculated for the equations of state of Ref. 1. ( astro-ph/0409602 )

## 2.  Basic characteristics of neutron stars with quark cores

The basic parameters of spherically symmetric static superdense stars are determined by numerical integration of the system of relativistic equations for the equilibrium of stars [2,3] supplemented by equations for the relativistic moment

Yerevan State University, Armenia

E-mail:  1) galaverdyan@ysu.am
         2) anharutr@ysu.am
         3) yuvartanyan@ysu.am

of inertia [4]. The following parameters of the stars are computed as functions of the central pressure $P_c$: radius $R$ ($P(R) = 0$), total mass $M$ ($M = 4\pi \int_0^R \rho r^2 dr$), rest mass $M_0$ ($M_0 = 4\pi m_0 \int_0^R r^2 n \exp(\lambda/2) dr$), proper mass $M_P$ ($M_P = 4\pi \int_0^R r^2 \rho \exp(\lambda/2) dr$), relativistic moment of inertia $I$, and red shift from the star's surface $Z_S$ ($Z_S = (1 - 2GM/Rc^2)^{-1/2} - 1$). Here $\rho$ is the total energy density ($\rho = m_0 n(1 + \varepsilon/m_0)$, $m_0 = M(^{56}Fe)/56$), $n$ is the baryon concentration, $\varepsilon$ is the average energy per baryon, and $\exp(\lambda)$ is the radial component of the metric tensor. The tables also list the radius $R_{core}$ and mass $M_{core} = 4\pi \int_0^{R_{core}} \rho r^2 dr$ of the quark core, as well as the radius $R_{Aen}$ and stored mass $M_{Aen} = 4\pi \int_0^{R_{Aen}} \rho r^2 dr$ corresponding to disappearance of degenerate neutrons, $\rho(R_{Aen}) = \rho_{drip} = 4.3 \cdot 10^{11}$ g/cm$^3$.

The computational results are grouped in accordance with the equations of state of the baryon component indicated by a number in the notation for the equations of state in the figures and tables. (1 corresponds to "HEA", 2 to "Bonn," and 3 to "BJ-V". See Ref. 2.) The quark components correspond to the letters in Table 1 of Ref. 1. Note that the dependence of the average energy e per baryon on the compressibility $1/n$ in the case of the baryon component is monotonic and for a given $1/n$ the energy is greatest for variant 3 (rigid equation of state) and minimal for variant 1 (soft equation of state). Variant 2 is close to 1 and lies between them (equation of state for intermediate rigidity). For the quark component, on the other hand, this dependence has a minimum and the softest variant, with the lowest value of $\varepsilon_{min}$ is variant $a$. Variants $b, c$, etc., follow and variants $g$ and $h$ are the most rigid, with the highest $\varepsilon_{min}$.

It is known [5,6] that the relativistic density discontinuity parameter $\lambda = \rho_Q / (\rho_N + P_0/c^2)$ ($\rho_N$ and $\rho_Q$ are the energy densities of the nucleon and quark phases, respectively, and $P_0$ is the transition pressure.) determines the dependence of the star's mass on the central pressure, $M(P_c)$. Thus, for $\lambda < 3/2$ the onset of the formation of the core of a quark phase in the center of the star corresponds to a discontinuity in the $M(P_c)$ curve without a change in the sign of the derivative. For $\lambda > 3/2$, on the other hand, the discontinuity is characterized by a local spike maximum in the rising branch of $M(P_c)$; i.e., at the threshold for formation of the core of the quark phase there is a descending branch, all of whose configurations are unstable (the so-called instability of configurations with low-mass cores).

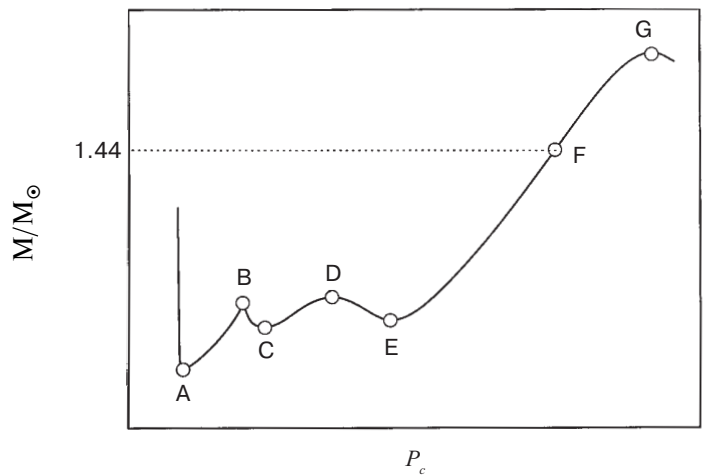

Fig. 1. Schematic illustration of the characteristic configurations whose parameters are tabulated in this paper.

TABLE 1. Main Integral and Structural Parameters of the Characteristic Configurations for Equations of State with a Density Discontinuity with $\lambda < 3/2$

| | $P_c$ | $\rho_c$ | $R_{core}$ | $M_{core}$ | $R_{Aen}$ | $m_{Aen}$ | $R$ | $M$ | $M_0$ | $I$ | $z_s$ |
|---|---|---|---|---|---|---|---|---|---|---|---|
| | MeV/fm$^3$ | $10^{14}$g/cm$^3$ | km | $M_\odot$ | km | $M_\odot$ | km | $M_\odot$ | $M_\odot$ | $M_\odot$ km$^2$ | |
| 1 | 2 | 3 | 4 | 5 | 6 | 7 | 8 | 9 | 10 | 11 | 12 |
| ES 1b | $P_0 = 5.979$ MeV/fm$^3$ | $\rho_N = 3.488 \cdot 10^{14}$ g/cm$^3$ | | | $\rho_Q = 4.859 \cdot 10^{14}$ g/cm$^3$ | | | $\lambda = 1.352$ | | | |
| A | 0.741 | 2.050 | 0 | 0 | 11.353 | 0.0762 | 245.007 | 0.0806 | 0.0811 | 9.439 | $4.9 \cdot 10^{-4}$ |
| B | 5.979 | 4.859 | 0 | 0 | 9.912 | 0.3183 | 12.221 | 0.3184 | 0.3273 | 8.776 | $4.1 \cdot 10^{-2}$ |
| F | 77.531 | 9.066 | 9.751 | 1.2091 | 11.345 | 1.4411 | 11.732 | 1.4411 | 1.6245 | 74.628 | $2.5 \cdot 10^{-1}$ |
| G | 321.975 | 22.801 | 9.775 | 1.7072 | 10.667 | 1.8315 | 10.876 | 1.8315 | 2.1431 | 90.304 | $4.1 \cdot 10^{-1}$ |
| ES 2b | $P_0 = 5.674$ MeV/fm$^3$ | $\rho_N = 3.523 \cdot 10^{14}$ g/cm$^3$ | | | $\rho_Q = 4.840 \cdot 10^{14}$ g/cm$^3$ | | | $\lambda = 1.336$ | | | |
| A | 0.741 | 2.014 | 0 | 0 | 11.265 | 0.0754 | 259.075 | 0.0798 | 0.0803 | 10.351 | $4.5 \cdot 10^{-4}$ |
| B | 5.674 | 4.840 | 0 | 0 | 9.933 | 0.3047 | 12.396 | 0.3048 | 0.3129 | 8.317 | $3.8 \cdot 10^{-2}$ |
| F | 77.531 | 9.066 | 9.786 | 1.2193 | 11.339 | 1.4400 | 11.726 | 1.4400 | 1.6230 | 74.466 | $2.5 \cdot 10^{-1}$ |
| G | 322.568 | 22.833 | 9.793 | 1.7130 | 10.662 | 1.8311 | 10.871 | 1.8311 | 2.1425 | 90.213 | $4.1 \cdot 10^{-1}$ |
| ES 2c | $P_0 = 14.109$ MeV/fm$^3$ | $\rho_N = 4.132 \cdot 10^{14}$ g/cm$^3$ | | | $\rho_Q = 5.263 \cdot 10^{14}$ g/cm$^3$ | | | $\lambda = 1.201$ | | | |
| A | 0.7407 | 2.014 | 0 | 0 | 11.265 | 0.0754 | 259.075 | 0.0798 | 0.0803 | 10.351 | $4.5 \cdot 10^{-4}$ |
| B | 14.109 | 4.132 | 0 | 0 | 11.130 | 0.6864 | 12.216 | 0.6864 | 0.7252 | 28.894 | $9.5 \cdot 10^{-2}$ |
| F | 67.284 | 8.308 | 8.723 | 0.8788 | 11.838 | 1.4400 | 12.272 | 1.4400 | 1.6119 | 79.270 | $2.4 \cdot 10^{-1}$ |
| G | 170.518 | 14.067 | 9.611 | 1.4377 | 11.577 | 1.8200 | 11.847 | 1.8200 | 2.1075 | 103.11 | $3.5 \cdot 10^{-1}$ |
| ES 3c | $P_0 = 6.015$ MeV/fm$^3$ | $\rho_N = 4.081 \cdot 10^{14}$ g/cm$^3$ | | | $\rho_Q = 4.790 \cdot 10^{14}$ g/cm$^3$ | | | $\lambda = 1.144$ | | | |
| A | 0.494 | 1.505 | 0 | 0 | 12.866 | 0.0877 | 197.845 | 0.0931 | 0.0934 | 9.379 | $6.9 \cdot 10^{-4}$ |
| B | 6.015 | 4.790 | 0 | 0 | 10.457 | 0.3070 | 13.211 | 0.3072 | 0.3133 | 8.676 | $3.6 \cdot 10^{-2}$ |
| F | 73.086 | 8.635 | 9.818 | 1.2003 | 11.537 | 1.4418 | 11.941 | 1.4419 | 1.6106 | 76.082 | $2.5 \cdot 10^{-1}$ |
| G | 170.518 | 14.067 | 10.225 | 1.6287 | 11.383 | 1.7989 | 11.646 | 1.7989 | 2.0760 | 99.515 | $3.5 \cdot 10^{-1}$ |
| ES 2d | $P_0 = 11.220$ MeV/fm$^3$ | $\rho_N = 3.970 \cdot 10^{14}$ g/cm$^3$ | | | $\rho_Q = 5.463 \cdot 10^{14}$ g/cm$^3$ | | | $\lambda = 1.310$ | | | |
| A | 0.741 | 2.008 | 0 | 0 | 11.264 | 0.0798 | 257.430 | 0.0798 | 0.0803 | 10.211 | $4.6 \cdot 10^{-4}$ |
| B | 11.220 | 3.970 | 0 | 0 | 10.759 | 0.5612 | 12.072 | 0.5612 | 0.5875 | 21.040 | $7.7 \cdot 10^{-2}$ |
| F | 82.864 | 9.566 | 9.004 | 1.0443 | 11.390 | 1.4400 | 11.782 | 1.4400 | 1.6181 | 73.480 | $2.5 \cdot 10^{-1}$ |
| G | 335.802 | 23.586 | 9.281 | 1.5942 | 10.599 | 1.8059 | 10.810 | 1.8059 | 2.1029 | 86.368 | $4.0 \cdot 10^{-1}$ |

TABLE 1. (continued)

| 1 | 2 | 3 | 4 | 5 | 6 | 7 | 8 | 9 | 10 | 11 | 12 |
|---|---|---|---|---|---|---|---|---|---|---|---|
| ES 3d  $P_0 = 5.291$ MeV/fm$^3$  $\rho_N = 3.870 \cdot 10^{14}$ g/cm$^3$  $\rho_Q = 5.117 \cdot 10^{14}$ g/cm$^3$  $\lambda = 291$ | | | | | | | | | | | |
| A | 0.494 | 1.505 | 0 | 0 | 12.866 | 0.0877 | 197.845 | 0.0931 | 0.0934 | 9.379 | 6.9·10$^{-4}$ |
| B | 5.291 | 5.117 | 0 | 0 | 10.474 | 0.2850 | 13.542 | 0.2852 | 0.2903 | 7.937 | 3.3·10$^{-2}$ |
| F | 87.7037 | 9.83888 | 9.667 | 1.2542 | 11.143 | 1.4402 | 11.512 | 1.4402 | 1.6158 | 71.301 | 2.6·10$^{-1}$ |
| G | 345.679 | 24.127 | 9.592 | 1.6946 | 10.440 | 1.7960 | 10.644 | 1.7960 | 2.0873 | 84.363 | 4.1·10$^{-1}$ |
| ES 2f  $P_0 = 20.272$ MeV/fm$^3$  $\rho_N = 4.485 \cdot 10^{14}$ g/cm$^3$  $\rho_Q = 6.103 \cdot 10^{14}$ g/cm$^3$  $\lambda = 1.259$ | | | | | | | | | | | |
| A | 0.741 | 2.014 | 0 | 0 | 11.265 | 0.0754 | 259.075 | 0.0798 | 0.0803 | 10.351 | 4.5·10$^{-4}$ |
| B | 20.272 | 6.103 | 0 | 0 | 11.75 | 0.9255 | 12.571 | 0.9256 | 0.9947 | 46.547 | 1.3·10$^{-1}$ |
| F | 76.889 | 9.407 | 7.854 | 0.7372 | 11.759 | 1.4400 | 12.185 | 1.4401 | 1.6101 | 77.138 | 2.4·10$^{-1}$ |
| G | 148.148 | 13.465 | 8.708 | 1.1548 | 11.471 | 1.6871 | 11.775 | 1.6871 | 1.9289 | 90.158 | 3.2·10$^{-1}$ |

The results of the calculations for the eleven equations of state are divided into two groups according to the two alternatives for the parameter $\lambda$ described above. For the seven equations of state corresponding to $\lambda < 3/2$, Table 1 lists the main integral parameters for the four characteristic configurations A, B, F and G (see Fig. 1). At the top are indicated the variant of the equation of state, the phase transition pressure $P_0$, and the energy density in the nucleon phase $\rho_N$ and in the quark phase $\rho_Q$ at the phase interface, as well as the value of the density discontinuity parameter $\lambda$. The softest quark equation of state (variant *a*) does not yield such a transition for any of the nucleon equations of state. For all seven variants the phase transition takes place at relatively high pressures. Thus, the lowest of these, the transition pressure for equation of state *3d* (the transition from a rigid nucleon equation of state to the quark equation of state *d*) is 5.29 MeV/fm$^3$ and the highest is for *2f*, with $P_0$ = 20.27 MeV/fm$^3$. For all seven variants the density of the nucleon phase, $\rho_N$, is higher than for the nuclear.

The A configurations refer to the point at which stability is lost on a plot of the mass as a function of central pressure, $M(P_c)$, at low masses. These configurations do not have a quark core. A large part of their mass is "Aen" matter; that is, it consists of degenerate electrons and neutrons, along with atomic nuclei with neutron excesses. The radius of this region is on the order of 10-11 km. Only 5% of the mass is in the "Ae" state, i.e., has the same composition as the matter in white dwarves. However, the radius of these configurations, which varies from 200 to 500 km, is mainly determined by the "Ae" matter. The gravitational packing coefficient $\alpha = (M_0 - M)/M$ for these configurations is in the range (3-6)×10$^{-3}$, or of the same order of magnitude as in white dwarves.

The B configurations refer to models of neutron stars for which the pressure at the center corresponds to the threshold for formation of a quark core. All the parameters of these models are also determined by the corresponding nucleon equations of state. Their masses vary over the range $(0.29 \div 0.93)M_\odot$ and their radii, over 12.1-13.5 km, while they consist almost entirely of "Aen" matter. For these models the gravitational packing coefficient α varies over $(1.8-8.5)\times 10^{-2}$ and the gravitational red shift from the surface, $Z_S$, over $(3.3-13)\times 10^{-2}$.

TABLE 2. Main Ontegral and Structural Parameters of the Characteristic Configurations for Equations of State with a Density Discontinuity with $\lambda > 3/2$

| | $P_c$ MeV/fm$^3$ | $\rho_c$ $10^{14}$g/cm$^3$ | $R_{core}$ km | $M_{core}$ $M_\odot$ | $R_{Aen}$ km | $m_{Aen}$ $M_\odot$ | $R$ km | $M$ $M_\odot$ | $M_0$ $M_\odot$ | $I$ $M_\odot$ | $z_s$ km$^2$ |
|---|---|---|---|---|---|---|---|---|---|---|---|
| ES 1a | $P_0 = 0.761$ MeV/fm$^3$ | | $\rho_N = 2.078 \cdot 10^{14}$ g/cm$^3$ | | | $\rho_Q = 4.467 \cdot 10^{14}$ g/cm$^3$ | | | $\lambda = 2.135$ | | | |
| A | 0.741 | 2.060 | 0 | 0 | 11.355 | 0.0762 | 248.331 | 0.0806 | 0.0811 | 9.704 | 4.8·10$^{-4}$ |
| B | 0.761 | 4.467 | 0 | 0 | 11.294 | 0.0768 | 197.209 | 0.0807 | 0.0812 | 6.25 | 6.0·10$^{-4}$ |
| C | 1.086 | 4.487 | 1.355 | 0.0023 | 11.255 | 0.0734 | 515.512 | 0.0798 | 0.0803 | 58.496 | 2.3·10$^{-4}$ |
| D | 1.195 | 4.493 | 1.562 | 0.0036 | 11.169 | 0.0723 | 702.629 | 0.0799 | 0.0804 | 145.042 | 1.7·10$^{-4}$ |
| E | 1.975 | 4.539 | 2.582 | 0.0162 | 10.016 | 0.0700 | 131.065 | 0.0723 | 0.0727 | 2.37 | 8.1·10$^{-4}$ |
| F | 74.469 | 8.719 | 10.494 | 1.4079 | 11.414 | 1.4391 | 11.053 | 1.4392 | 1.6359 | 74.795 | 2.6·10$^{-1}$ |
| G | 296.296 | 21.054 | 10.332 | 1.8453 | 10.854 | 1.8626 | 10.653 | 1.8626 | 2.2039 | 95.491 | 4.2·10$^{-1}$ |
| ES 2a | $P_0 = 0.758$ MeV/fm$^3$ | | $\rho_N = 2.028 \cdot 10^{14}$ g/cm$^3$ | | | $\rho_Q = 4.466 \cdot 10^{14}$ g/cm$^3$ | | | $\lambda = 2.188$ | | | |
| A | 0.741 | 2.014 | 0 | 0 | 11.265 | 0.0754 | 259.075 | 0.0798 | 0.0803 | 10.351 | 4.5·10$^{-4}$ |
| B | 0.758 | 4.466 | 0 | 0 | 11.213 | 0.0760 | 207.209 | 0.0799 | 0.0804 | 6.652 | 5.7·10$^{-4}$ |
| C | 0.938 | 4.478 | 1.012 | 0.0030 | 11.246 | 0.0740 | 386.021 | 0.0795 | 0.0800 | 26.41 | 3.0·10$^{-4}$ |
| D | 1.294 | 4.498 | 1.734 | 0.0049 | 10.995 | 0.0705 | 1304.496 | 0.0820 | 0.0825 | 1062.998 | 9.3·10$^{-5}$ |
| E | 1.975 | 4.539 | 2.586 | 0.0163 | 9.948 | 0.0694 | 133.876 | 0.0717 | 0.0721 | 2.376 | 7.9·10$^{-4}$ |
| F | 74.568 | 8.724 | 10.496 | 1.4089 | 11.05 | 1.4399 | 11.411 | 1.4400 | 1.6369 | 74.853 | 2.6·10$^{-1}$ |
| G | 320.988 | 22.410 | 10.264 | 1.8470 | 10.575 | 1.8635 | 10.772 | 1.8635 | 2.2052 | 94.121 | 4.3·10$^{-1}$ |
| ES 3a | $P_0 = 0.199$ MeV/fm$^3$ | | $\rho_N = 0.864 \cdot 10^{14}$ g/cm$^3$ | | | $\rho_Q = 4.433 \cdot 10^{14}$ g/cm$^3$ | | | $\lambda = 5.112$ | | | |
| A | 0.035 | 0.277 | 0 | 0 | 18.168 | 0.0523 | 893.565 | 0.6101 | 0.6157 | 2.449·10$^4$ | 1.0·10$^{-3}$ |
| B | 0.199 | 4.433 | 0 | 0 | 14.583 | 0.0723 | 2214.655 | 0.6386 | 0.6444 | 2.363·10$^5$ | 4.3·10$^{-4}$ |
| C | 0.395 | 4.445 | 1.066 | 0.0011 | 14.897 | 0.0665 | 1708.299 | 0.6360 | 0.6418 | 1.210·10$^5$ | 5.5·10$^{-4}$ |
| D | 2.123 | 4.547 | 3.253 | 0.0324 | 8.286 | 0.0459 | 2696.469 | 0.8200 | 0.8277 | 6.315·10$^4$ | 4.5·10$^{-4}$ |
| E | 2.395 | 4.563 | 3.461 | 0.0390 | 7.755 | 0.0514 | 361.656 | 0.0534 | 0.0535 | 9.204 | 2.2·10$^{-4}$ |
| F | 74.568 | 8.724 | 10.567 | 1.4309 | 10.935 | 1.4390 | 11.286 | 1.4390 | 1.6356 | 74.715 | 2.7·10$^{-1}$ |
| G | 316.049 | 22.139 | 10.318 | 1.9000 | 10.527 | 1.8632 | 10.72 | 1.8632 | 2.2047 | 94.352 | 4.3·10$^{-1}$ |
| ES 3b | $P_0 = 0.796$ MeV/fm$^3$ | | $\rho_N = 1.729 \cdot 10^{14}$ g/cm$^3$ | | | $\rho_Q = 4.543 \cdot 10^{14}$ g/cm$^3$ | | | $\lambda = 2.606$ | | | |
| A | 0.494 | 1.505 | 0 | 0 | 12.866 | 0.0877 | 197.845 | 0.0931 | 0.0934 | 9.379 | 6.9·10$^{-4}$ |
| B | 0.796 | 4.543 | 0 | 0 | 11.952 | 0.1025 | 48.493 | 0.1042 | 0.1047 | 3.173 | 3.2·10$^{-3}$ |
| C | 2.222 | 4.631 | 2.742 | 0.0198 | 10.751 | 0.0870 | 53.782 | 0.0885 | 0.0887 | 1.939 | 2.4·10$^{-3}$ |
| F | 79.506 | 9.180 | 10.386 | 1.4079 | 11.003 | 1.4393 | 11.360 | 1.4393 | 1.6209 | 73.282 | 2.6·10$^{-1}$ |
| G | 325.926 | 23.019 | 10.113 | 1.8093 | 10.467 | 1.8262 | 10.666 | 1.8262 | 2.1348 | 89.237 | 4.2·10$^{-1}$ |

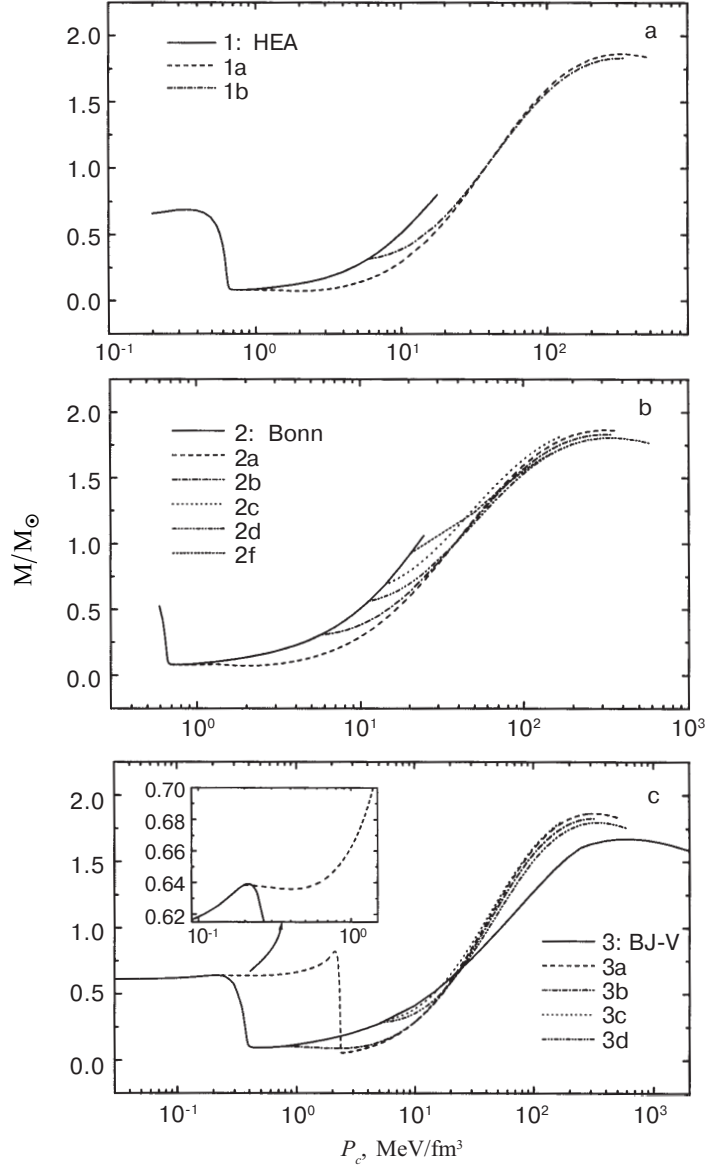

Fig. 2. The gravitational mass $M$ as a function of the central pressure $P_c$ for sets of equations of state with variants "HEA" (2a), "Bonn" (2b), and "BJ-V" (2c). The smooth curves correspond to models of neutron stars without a quark core and with the given variant for the nucleon matter. The magnified insert shows the region of the phase transition for the equation of state 3a.

In all seven variants of the equation of state for the observationally most accurately determined neutron star masses $M = 1.44 M_\odot$ [7], we list the computational results for the other parameters (configuration F). For these models the value of the gravitational packing coefficient a reaches 0.1, the ratio $M_{core}/M$ of the masses of the quark core to the entire mass of the star ranges over 0.51-0.87, and the ratio $R_{core}/R$ of the radius of the quark core to the star's radius, over 0.64-0.84. The gravitational red shift from the surface, $Z_S$, ranges over 0.24-0.26.

The G configurations refer to superdense stars with a maximum mass for which stability is lost. The maximum mass $M_{max}$ lies within the limits $(1.69 \div 1.83)M_\odot$, the gravitational packing coefficient α, within 0.14-0.17, the ratio $M_{core}/M$ of the masses of the quark core to the star's mass, within 0.68-0.94, and the ratio $R_{core}/R$ of the radii, within 0.74-0.9. The gravitational red shift from the surface, $Z_S$, for these models lies between 0.32-0.41.

Table 2 lists the integral parameters of the characteristic configurations for the four equations of state 1a, 2a, 3a, and 3b with λ > 3/2. A phase transition of this type is obtained for all the nucleon equations of state in the case of the

TABLE 3. Results of Intermediate Integrations for the Equation of State 3a. (See text.) The Characteristic Configurations are Indicated

| | $P_c$ MeV/fm³ | $\rho_c$ 10¹⁴g/cm³ | $R_{core}$ km | $M_{core}$ $M_\odot$ | $R_{Aen}$ km | $m_{Aen}$ $M_\odot$ | $R$ km | $M$ $M_\odot$ | $M_0$ $M_\odot$ | $I$ $M_\odot$ | $z_s$ km² |
|---|---|---|---|---|---|---|---|---|---|---|---|
| ES 3a | $P_0 = 0.199$ MeV/fm³ | | | | $\rho_N = 0.864 \cdot 10^{14}$ g/cm³ | | $\rho_Q = 4.433 \cdot 10^{14}$ g/cm³ | | | λ = 5.112 | |
| 1 | 2 | 3 | 4 | 5 | 6 | 7 | 8 | 9 | 10 | 11 | 12 |
| | 0.0247 | 0.2219 | 0 | 0 | 18.871 | 0.0491 | 812.959 | 0.6109 | 0.6165 | 2.005·10⁴ | 1.1·10⁻³ |
| | 0.0296 | 0.2492 | 0 | 0 | 18.491 | 0.0508 | 853.993 | 0.6103 | 0.6159 | 2.223·10⁴ | 1.1·10⁻³ |
| A | 0.0346 | 0.2765 | 0 | 0 | 18.168 | 0.0523 | 893.565 | 0.6101 | 0.6157 | 2.449·10⁴ | 1.0·10⁻³ |
| | 0.0444 | 0.3275 | 0 | 0 | 17.647 | 0.0547 | 968.035 | 0.6103 | 0.6159 | 2.921·10⁴ | 9.3·10⁻⁴ |
| | 0.0494 | 0.3513 | 0 | 0 | 17.427 | 0.0558 | 1004.338 | 0.6107 | 0.6162 | 3.174·10⁴ | 9.0·10⁻⁴ |
| | 0.0741 | 0.4569 | 0 | 0 | 16.583 | 0.0601 | 1182.169 | 0.6138 | 0.6193 | 4.646·10⁴ | 7.7·10⁻⁴ |
| | 0.0988 | 0.5520 | 0 | 0 | 15.995 | 0.0633 | 1360.636 | 0.6181 | 0.6236 | 6.557·10⁴ | 6.7·10⁻⁴ |
| | 0.1235 | 0.6383 | 0 | 0 | 15.542 | 0.0660 | 1546.957 | 0.6232 | 0.6288 | 9.090·10⁴ | 5.9·10⁻⁴ |
| | 0.1481 | 0.7167 | 0 | 0 | 15.174 | 0.0683 | 1747.245 | 0.6288 | 0.6345 | 1.251·10⁵ | 5.3·10⁻⁴ |
| | 0.1728 | 0.7915 | 0 | 0 | 14.867 | 0.0703 | 1961.386 | 0.6343 | 0.6400 | 1.705·10⁵ | 4.8·10⁻⁴ |
| | 0.1975 | 0.8584 | 0 | 0 | 14.602 | 0.0722 | 2195.380 | 0.6384 | 0.6442 | 2.309·10⁵ | 4.3·10⁻⁴ |
| B | 0.1995 | 0.8637 | 0 | 0 | 14.583 | 0.0723 | 2214.655 | 0.6386 | 0.6444 | 2.363·10⁵ | 4.3·10⁻⁴ |
| | | 4.4330 | | | | | | | | | |
| | 0.2025 | 4.4339 | 0.133 | 2.2·10⁻⁶ | 14.591 | 0.0723 | 2206.151 | 0.6385 | 0.6443 | 2.339·10⁵ | 4.3·10⁻⁴ |
| | 0.2469 | 4.4365 | 0.527 | 1.4·10⁻⁴ | 14.689 | 0.0711 | 2077.946 | 0.6378 | 0.6436 | 1.998·10⁵ | 4.5·10⁻⁴ |
| | 0.2963 | 4.4392 | 0.751 | 3.9·10⁻⁴ | 14.776 | 0.0697 | 1943.711 | 0.6370 | 0.6428 | 1.680·10⁵ | 4.8·10⁻⁴ |
| | 0.3704 | 4.4436 | 0.997 | 9.2·10⁻⁴ | 14.874 | 0.0673 | 1762.334 | 0.6361 | 0.6419 | 1.308·10⁵ | 5.3·10⁻⁴ |
| C | 0.3951 | 4.4453 | 1.066 | 0.001 | 14.897 | 0.0665 | 1708.299 | 0.6360 | 0.6418 | 1.210·10⁵ | 5.5·10⁻⁴ |
| | 0.4938 | 4.4515 | 1.306 | 0.002 | 14.950 | 0.0630 | 1517.060 | 0.6367 | 0.6425 | 9.066·10⁴ | 6.2·10⁻⁴ |

TABLE 3. (continued)

| 1 | 2 | 3 | 4 | 5 | 6 | 7 | 8 | 9 | 10 | 11 | 12 |
|---|---|---|---|---|---|---|---|---|---|---|---|
|   | 0.6173 | 4.4585 | 1.552 | 0.004 | 14.904 | 0.0586 | 1334.963 | 0.6401 | 0.6459 | $6.757 \cdot 10^4$ | $7.1 \cdot 10^{-4}$ |
|   | 0.8642 | 4.4735 | 1.950 | 0.007 | 14.384 | 0.0499 | 1108.907 | 0.6528 | 0.6588 | $4.603 \cdot 10^4$ | $8.7 \cdot 10^{-4}$ |
|   | 1.1111 | 4.4876 | 2.275 | 0.011 | 13.231 | 0.0431 | 1022.746 | 0.6711 | 0.6773 | $4.090 \cdot 10^4$ | $9.7 \cdot 10^{-4}$ |
|   | 1.6049 | 4.5166 | 2.803 | 0.021 | 10.222 | 0.0393 | 1227.599 | 0.7244 | 0.7312 | $7.278 \cdot 10^4$ | $8.7 \cdot 10^{-4}$ |
|   | 1.8519 | 4.5316 | 3.027 | 0.026 | 9.117 | 0.0416 | 1627.587 | 0.7653 | 0.7725 | $1.572 \cdot 10^5$ | $6.9 \cdot 10^{-4}$ |
|   | 2.0247 | 4.5422 | 3.173 | 0.030 | 8.549 | 0.0442 | 2181.119 | 0.8084 | 0.8160 | $3.582 \cdot 10^5$ | $5.5 \cdot 10^{-4}$ |
| D | 2.1235 | 4.5475 | 3.253 | 0.032 | 8.286 | 0.0459 | 2696.469 | 0.8200 | 0.8277 | $6.315 \cdot 10^5$ | $4.5 \cdot 10^{-4}$ |
|   | 2.2716 | 4.5563 | 3.368 | 0.036 | 7.972 | 0.0488 | 4524.922 | 0.6600 | 0.6662 | $1.764 \cdot 10^6$ | $2.2 \cdot 10^{-4}$ |
|   | 2.3704 | 4.5624 | 3.443 | 0.038 | 7.798 | 0.0508 | 969.389 | 0.0548 | 0.0549 | $2.028 \cdot 10^2$ | $8.3 \cdot 10^{-5}$ |
| E | 2.395 | 4.563 | 3.461 | 0.039 | 7.755 | 0.0514 | 361.656 | 0.0534 | 0.0535 | 9.204 | $2.2 \cdot 10^4$ |
|   | 2.4691 | 4.5677 | 3.515 | 0.041 | 7.650 | 0.0530 | 122.953 | 0.0541 | 0.0543 | 0.918 | $6.5 \cdot 10^4$ |
|   | 3.4568 | 4.6267 | 4.149 | 0.068 | 6.972 | 0.0782 | 16.482 | 0.0784 | 0.0789 | 0.664 | 0.007 |
|   | 5.9259 | 4.7720 | 5.308 | 0.140 | 7.055 | 0.1535 | 9.911 | 0.1536 | 0.1560 | 1.903 | 0.024 |
|   | 7.4074 | 4.8591 | 5.834 | 0.190 | 7.310 | 0.2020 | 9.405 | 0.2021 | 0.2063 | 2.986 | 0.033 |
|   | 8.8889 | 4.9463 | 6.280 | 0.240 | 7.576 | 0.2512 | 9.266 | 0.2513 | 0.2579 | 4.280 | 0.043 |
|   | 10.3704 | 5.0335 | 6.665 | 0.290 | 7.831 | 0.3003 | 9.268 | 0.3004 | 0.3098 | 5.756 | 0.052 |
|   | 16.7901 | 5.4085 | 7.877 | 0.490 | 8.732 | 0.5030 | 9.665 | 0.5031 | 0.5287 | 13.57 | 0.087 |
|   | 19.7531 | 5.5811 | 8.275 | 0.580 | 9.049 | 0.5886 | 9.871 | 0.5886 | 0.6232 | 17.61 | 0.10 |
|   | 24.6914 | 5.8681 | 8.797 | 0.710 | 9.473 | 0.7187 | 10.170 | 0.7188 | 0.7694 | 24.52 | 0.12 |
|   | 29.6296 | 6.1543 | 9.194 | 0.820 | 9.803 | 0.8341 | 10.417 | 0.8341 | 0.9015 | 31.34 | 0.14 |
|   | 34.5679 | 6.4386 | 9.504 | 0.930 | 10.059 | 0.9361 | 10.614 | 0.9362 | 1.0203 | 37.86 | 0.16 |
|   | 39.5062 | 6.7230 | 9.749 | 1.000 | 10.265 | 1.0264 | 10.774 | 1.0264 | 1.1270 | 43.97 | 0.18 |
|   | 64.1975 | 8.1352 | 10.430 | 1.300 | 10.827 | 1.3483 | 11.208 | 1.3483 | 1.5207 | 67.73 | 0.24 |
| F | 74.568 | 8.724 | 10.567 | 1.4309 | 10.935 | 1.4390 | 11.286 | 1.4390 | 1.6356 | 74.71 | 0.27 |
|   | 75.0617 | 8.7524 | 10.572 | 1.400 | 10.938 | 1.4429 | 11.288 | 1.4429 | 1.6405 | 75.01 | 0.27 |
|   | 83.9506 | 9.2552 | 10.650 | 1.500 | 10.997 | 1.5060 | 11.327 | 1.5060 | 1.7218 | 7.983 | 0.28 |
|   | 148.1481 | 12.8588 | 10.759 | 1.700 | 11.027 | 1.7508 | 11.280 | 1.7508 | 2.0475 | 9.641 | 0.36 |
|   | 296.2963 | 21.0539 | 10.374 | 1.900 | 10.586 | 1.8623 | 10.783 | 1.8623 | 2.2034 | 9.544 | 0.43 |
|   | 311.1111 | 21.8674 | 10.332 | 1.900 | 10.541 | 1.8631 | 10.735 | 1.8631 | 2.2045 | 9.463 | 0.43 |
| G | 316.0494 | 22.1386 | 10.318 | 1.900 | 10.527 | 1.8632 | 10.720 | 1.8632 | 2.2047 | 9.435 | 0.43 |
|   | 330.8642 | 22.9512 | 10.277 | 1.900 | 10.482 | 1.8630 | 10.673 | 1.8630 | 2.2044 | 9.351 | 0.44 |
|   | 340.7407 | 23.4927 | 10.250 | 1.900 | 10.454 | 1.8625 | 10.644 | 1.8626 | 2.2037 | 9.293 | 0.44 |

softest quark equation of state (variant *a*), for which the average energy per baryon is lowest at the minimum point: $\varepsilon_{min} = 10.44$ MeV [1]. For the second softest quark equation of state, *b*, with $\varepsilon_{min} = 20.71$ MeV this transition only occurs in the case of the rigid nucleon equation of state 3. The phase transition pressure is much lower for all four equations of state than for the cases with λ<3/2 examined above. The density of the nucleon phase at the transition point is lower than that of the nuclear. Variant 3a, for which the transition pressure and density of the nucleon phase are the lowest at $P_0 = 0.199$ MeV/fm$^3$ and $\rho_N = 0.861 \cdot 10^{14}$ g/cm$^3$, stands out particularly in this case. For this equation of state $\lambda$ is highest, with $\lambda = 5.112$.

As noted above, when $\lambda > 3/2$ a spike discontinuity develops in the rising branch of the $M(P_c)$ curve at the phase transition point. Unlike in Ref. 8, in our models this transition takes place in the low mass region for the equations of state 1a, 2a, and 3b. Only case 3b, for which the transition pressure is the highest for the four equations of state with $\lambda > 3/2$, has just the characteristic spike discontinuity. In the cases of the equations of state 1a and 2a, on the other hand, after this discontinuity a small additional maximum appears. Models corresponding to the equation of state 2a were first examined in Refs. 9 and 10. These models are interesting in that the presence of a second local maximum ensures the possibility of a new branch of stable equilibrium configurations - neutron stars with strange quark cores and radii exceeding 1000 km.

The equation of state 3a is of special interest in this regard. Thus, while the phase transition with an additional mass peak lies in the low mass region for the equations of state 1a and 2a, for the equation of state model 3a the phase transition is shifted into a region of intermediate mass white dwarves ($M \sim 0.6 M_\odot$) lying beyond the Chandrasekhar limit. Note that the nucleon equation of state 3, which is based on the data of Ref. 11 in this range of densities, yields a small maximum beyond the Chandrasekhar limit even in the case where no transition to the quark phase is seen (smooth curve in Fig. 2c). This can be seen clearly in Fig. 1 of Ref. 11, as well. Unfortunately, there are no tabulated data relevant to this region in that paper. Thus, we have studied this region in some detail, including the case without a transition to a quark phase. The results are shown in Table 3. The mass as a function of the central density, $M(\rho_c)$, is characterized by the condition $dM/d\rho_c < 0$ beyond the Chandrasekhar limit; this corresponds to the unstable white dwarves. The sign of the derivative does not change even after the central density $\rho_c$ becomes greater than $\rho_{drip} = 4.3 \times 10^{11}$ g/cm$^3$, at which a degenerate neutron gas is formed - and the stellar material changes to an "Aen" state. In the case of the nucleon equations of state, here for $M \approx (0.7 \div 0.6) M_\odot$ the decrease in the $M(\rho_c)$ curve initially slows down and then, at some $\rho_c$, it drops sharply - low-mass neutron stars are formed and the sign of the derivative changes: $dM/d\rho_c$ becomes positive, which corresponds to the stable branch of the neutron stars. In the case of the equation of state 3, however, as Table 3 shows, for central densities $3.5 \cdot 10^{13}$ g/cm$^3 \leq \rho_c \leq 8.6 \cdot 10^{13}$ g/cm$^3$, $dM/d\rho_c > 0$: an additional region of stable white dwarves develops with intermediate masses and small cores containing "Aen" matter. For the limiting configuration we have: $M_{max} = 0.638 M_\odot$, $R_{max} = 2195$ km, gravitational packing coefficient $\alpha = 9 \cdot 10^{-3}$, $M_{Aen} = 0.072 M_\odot$, and $R_{Aen} = 14.6$ km. Note that in this additional stable branch, as opposed to the case of ordinary white dwarves and neutron stars, the radius also increases as the density rises.

Table 2 shows that, in the case where a quark phase is formed with the equation of state 3a, the spike discontinuity (region AB) and the additional local maximum for neutron stars with a quark core (region CD) that follows it both shift into this region of additional stability for intermediate mass white dwarves.

We now consider Table 2. Here, besides the four configurations A, B, F, and G, which have the same significance as in Table 1 for the equations of state 1a, 2a, and 3a, we list the parameters of the three configurations C, D, and E, which describe a local maximum that appears after the spike discontinuity at the phase transition to the quark state, as well those for the equation of state 3b, configuration C, which describes a minimum that develops after the spike discontinuity and is absent in Table 1. Region AB applies to stable superdense configurations without a quark core and CD, to configurations with a small quark core.

Configurations A, B, C, and D are extremely close in mass, which may raise doubt as to the regularity of the data for them. Thus, integrals were taken for more than one hundred intermediate configurations for each equation of state; the results of these are shown separately in Figs. 2c, 3a, 3b, and 3c separately for this segment (insets).

On segment AB, which corresponds to low mass stable neutron stars without a quark core ($dM/d\rho_c > 0$) for equations of state 1a, 2a, and 3b, the radius decreases with increasing mass, as should happen for ordinary stable neutron stars. This behavior is not observed only with the equation of state 3a, for which the radius also increases with increasing mass on this branch. In the case of equation of state 3a, as opposed to the first three, the ratios $M_{Aen}/M$ and $R_{Aen}/R$ for configuration B take values of 0.11 and $6.6 \times 10^{-3}$, respectively; that is, the overwhelming bulk of the material is in the "Ae" state, as in white dwarves.

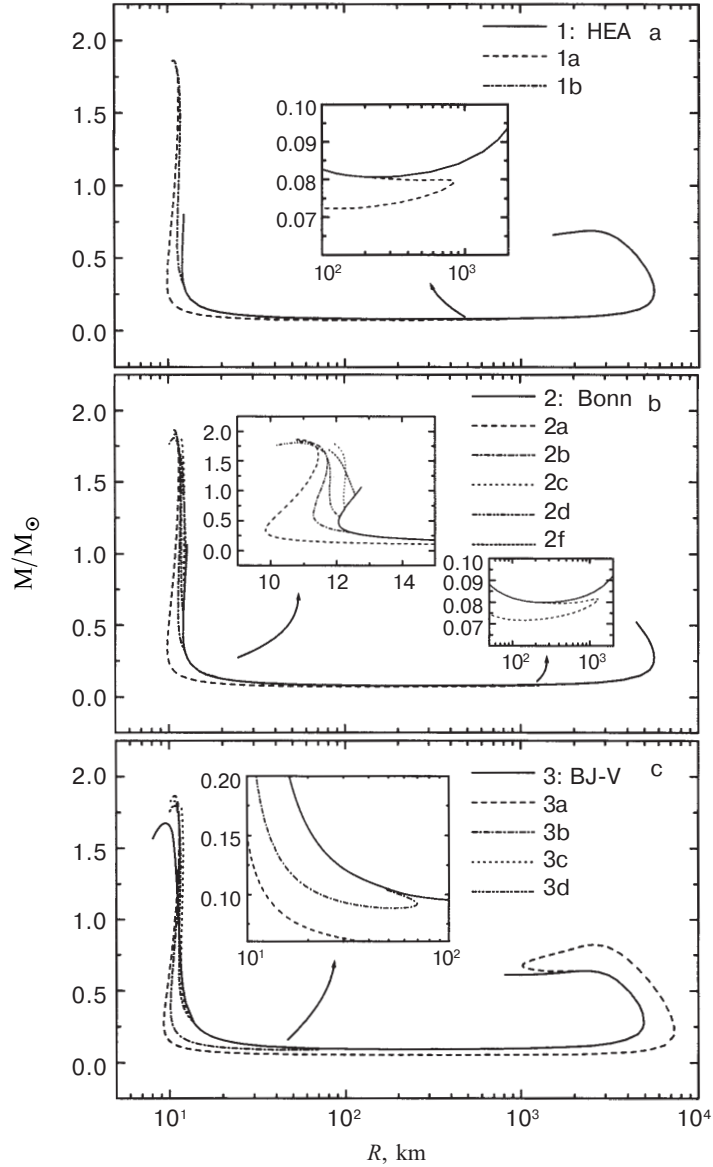

Fig. 3. The mass $M$ as a function of radius $R$. The magnified inserts show the regions with an additional local maximum in the mass of the neutron stars with strange quark cores (See text.), and in Fig. 3b the insert in the upper left corner also shows the region of the phase transition for the entire set of equations of state.

Segment CD corresponds to stable neutron stars with a small quark core whose mass is as much as a few percent of the star's entire mass. On this branch for all three equations of state 1a, 2a, and 3a, for which such a region exists, the radius increases with the mass. As on AB, the gravitational packing coefficient $a$ on segment CD has the same values as for massive white dwarves. In the future we also plan to study the stability of these segments using Chandrasekhar's standard variation method [12].

While $M_B > M_D$ for the equation of state 1a, the opposite, $M_D > M_B$, holds for 2a and 2a. This means that for the latter two equations of state, there may be a transition with accretion from configuration B to a denser state as a result of two discontinuous transitions: first to the configuration of branch CD and then to the main rising branch of stable stars. Here there would be a double release of energy in the form of two sequential supernova explosions [9,10].

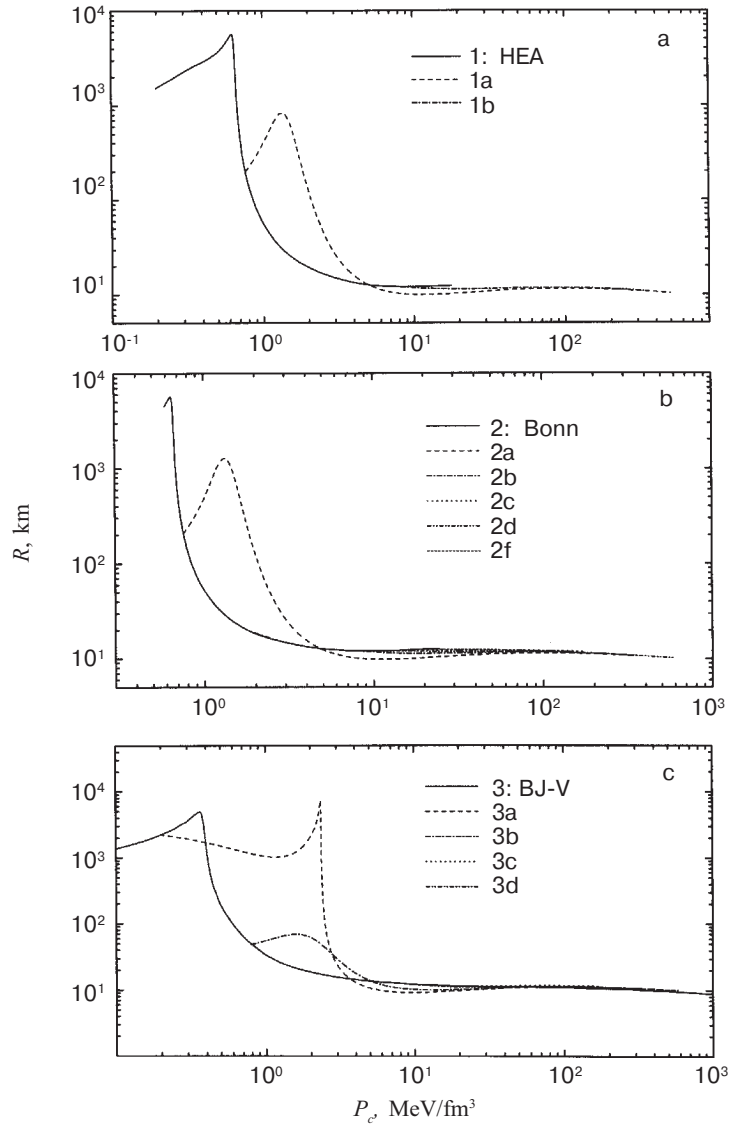

Fig. 4. The radius $R$ as a function of central pressure $P_c$. (Notation as in Figs. 2 and 3).

For all the equations of state, $R_B < R_D$, which are equal, respectively to 2700 (3a), 1300 (2a), and 700 km (1a).

Table 2 shows that for the F configurations, which describe a fixed mass $M = 1.44 M_\odot$, when $\lambda > 3/2$ the contribution of the quark core to the overall mass is as high as 978%, while when $\lambda < 3/2$ this fraction is no more than 83%. This fact is explained by the transition to a quark phase at lower densities when $\lambda > 3/2$.

For the models with the maximum mass (configuration G) when $\lambda > 3/2$ the contribution of the quark core reaches 99%. The fact that the parameters of this configuration are almost identical for all three equations of state 1a, 2a, and 3a shows that in the region of the maximum mass the quark component of the equation of state, which is the same "*a*", predominates.

Figures 2a, 2b, and 2c show the dependence of the gravitational mass M on the central pressure $P_c$ for sets of equations of states with the variants "HEA," "Bonn," and "BJ-V," respectively. In all the graphs the continuous curves correspond to neutron stars without a quark core as calculated using the "HEA," "Bonn," and "BJ-V" equations of state.

The mass *M* as a function of radius *R* and the radius *R* as a function of the pressure $P_c$ for the sets of equations of state with nucleon components "HEA," "Bonn," and "BJ-V" are plotted in Figs. 3a-c and 4a-c, respectively.

3. Conclusion

A large set of equations of state of superdense matter with a quark phase transition has been used to calculate a series of models for layered neutron stars with a strange quark core. A comparative analysis has been made of the functional dependences of their integral parameters and structural characteristics on the form of the equation of state.

For some models with a transition parameter $\lambda > 3/2$ a new branch of stable configurations with a small quark core ($M_{core} \sim 0.004 \div 0.03 M_\odot$) has been found on plots of the mass as a function of the central pressure. For some of the models $M_{max} \sim 0.08 M_\odot$ at the maximum and for the others $M_{max} \sim 0.82 M_\odot$. The layered neutron stars which lie on this branch are also characterized by unusually large radii (ranging from $R \sim 1000$ km to $R \sim 2500$ km for different models).

This work was supported by the Armenian National Science and Education Foundation (ANSEF grant No. PS 140) and was carried out in the framework of topic #0842 financed by the Ministry of Education and Science of the Armenian Republic.